# Bright single photon sources in lateral silicon carbide light emitting diodes

Matthias Widmann[1*], Matthias Niethammer[1], Takahiro Makino[2], Torsten Rendler[1], Stefan Lasse[1], Takeshi Ohshima[2], Jawad Ul Hassan[3], Nguyen Tien Son[3], Sang-Yun Lee[4†] & Jörg Wrachtrup[1]

[1]3. Physikalisches Institut and Research Center SCOPE, University Stuttgart, Pfaffenwaldring 57, 70569 Stuttgart, Germany

[2]National Institutes for Quantum and Radiological Science and Technology, Takasaki, Gunma 370-1292, Japan

[3]Department of Physics, Chemistry and Biology, Linköping University, SE-58183 Linköping, Sweden

[4]Center for Quantum Information, Korea Institute of Science and Technology, Seoul, 02792, Republic of Korea

*m.widmann@physik.uni-stuttgart.de

†sangyun.lee@kist.re.kr

Single-photon emitting devices have been identified as an important building block for applications in quantum information and quantum communication. They allow to transduce and collect quantum information over a long distance via photons as so called flying qubits. In addition, substrates like silicon carbide provides an excellent material platform for electronic devices. In this work we combine these two features and show that one can drive single photon emitters within a silicon carbide p-i-n-diode.

To achieve this, we specifically designed a lateral oriented diode. We find a variety of new color centers emitting non-classical lights in VIS and NIR range. One type of emitter can be electrically excited, demonstrating that silicon carbide can act as an ideal platform for electrically controllable single photon sources.

Many color centers in wide-bandgap semiconductors are considered as prominent quantum light sources owing to their stability, brightness, and spin-dependent properties[1–5]. Despite the wealth color centers found, there is an ongoing demand in e.g. quantum communication, information processing and sensing to explore new candidate defects. Since color centers are point defects, whose excited states can decay radiatively, searching natural materials or artificially grown but impure crystals may allow for identifying new color centers[6,7]. In addition, artificially adjusted material properties can lead to the creation of optical and spin active color centers. For example, by gradually adjusting both donor and acceptor impurity concentrations across a chemical vapor deposition (CVD) grown diamond crystal, the neutral silicon vacancy center could be reproducible created *via* charge state alignment[8]. While most color centers are driven optically, electric excitation provides another method to drive optical transitions. Furthermore it provides an efficient way to integrate quantum light sources into compact electronic devices[9], as there is no need for an additional bulky laser system.

In this report, we fabricate lateral silicon carbide (SiC) diodes by ion implantation of donor and acceptor impurities forming 20 µm wide p-i-n junctions in a nearly two-dimensional layer. This structure provides easy optical access to three distinctive layers, in which the Fermi-level is modulated over a wide energy range. In addition, because efficient carrier injection is possible, our structure provides an opportunity to look for electrically driven color centers. Similar lateral p-i-n junctions on diamond surface were reported, to study charge state conversion and electrical excitation of the nitrogen-vacancy centers in diamond[10]. However, the inherent properties of diamond disallow efficient doping, e.g. mismatch of depth of doped layers due to adverse production of thick graphite layers on the surface[10,11]. SiC is known to be an industry friendly platform, whose doping by ion implantation is very efficient[12]. Furthermore, ion implantation allows localized doping in the sub-microscopic scale, making it a highly necessary technique for scalable quantum device fabrication[10,13]. Besides the well-studied color centers in silicon carbide such as the silicon vacancy and divacancy,[4,14–16] several bright new color centers have been identified in bulk SiC substrate and electronic and optoelectronic devices[13,17–20]. Among recent reports, a p-n junction was used to electrically excite the silicon antisite defect[13]. While p-n junctions allows efficient carrier injection, incorporation of an intrinsic layer as in our study is beneficial, as it allows gradual Fermi-level tuning in a wide area. It may enhance the chance to find new color centers, with enhancing brightness in SiC as suggested recently[21]. Here, we demonstrate varying optical properties of various color center, hitherto unknown single photon sources in 4H-SiC lateral p-i-n diodes including one which can be electrically pumped. We investigate their fluorescence spectra decay dynamics as a function of optical or electrical pump power.

A large part of the fabricated lateral p-i-n devices is presented in the optical microscope image in Fig. 1(a). The p-i-n diodes were fabricated on a vertical stack of epilayers that was grown by CVD with the following structure: $n^+$ 4H-SiC substrate/semi-insulating (SI) layer (10 μm thick)/$n^-$ (17 μm thick). SI layer was grown through in-situ vanadium doping during growth using vanadium tetrachloride (VTC) in bubbler configuration[22] A high concentration of vanadium is known to produce extended defects in the epilayer. The flow rate of VTC was optimized to achieve defect free epilayer with SI properties (resistivity >$10^5$ Ωcm). The $n^-$ layer was grown under Si-limited condition to ensure low n-type residual doping of nitrogen. The free carrier concentration of the undoped $n^-$ layer at room temperature is in the range mid-to-high $10^{13}$ cm$^{-3}$ as determined from the capacitance versus voltage measurements. SI and

n⁻ layers were grown in a single growth run to avoid the formation of new defects at the interface.

N- and p-layers were created using phosphorus (P) and aluminum (Al) implantation, respectively at 800 °C. Three-fold implantation of $P^+$ (200, 140 and 80 keV) and $Al^+$ (110, 75 and 50 keV) was carried out to obtain donor and acceptor concentrations of $2\times10^{20}$ and $5\times10^{19}$ cm$^{-3}$, respectively. Implantation depth have been designed to 300 nm for n- and 200 nm for p-layers. After ion implantation, samples were annealed at 1800 °C for 5 min in an Ar atmosphere. To avoid surface degradation[23], the sample surface has been covered with a carbon film during high temperature annealing. A 45 nm thick field oxide layer was formed on the SiC surface using pyrogenic oxidation ($H_2:O_2$ = 1:1) at 1100 °C. To get electronic access to our device, metal electrodes on n- and p-layers have been formed by Al evaporation and lithography with liftoff technique. Each of these diodes can be individually addressed via the created top Al electrodes. While each p-doped region is covered by its own Al electrode, the n-layers are connected all together with a common Al electrode. In Fig. 1(b), a schematic of the p-i-n diodes with 20 μm width is illustrated. The n-layer is separated from the p-layer by an intrinsic layer (10 μm long). The top electrodes are shown in grey and are placed on top of the n- and p-layers, respectively. Note that this design is also used for a similar study reporting new color centers at the oxidized silicon carbide surface[20]. A typical bias-current measurement curve is plotted in Fig. 1(c), illustrating the typical diode behavior. The device is mounted on a home-built confocal microscope[4], equipped with single photon counting modules (SPCMs) and an high NA oil objective (NA=1.35) to monitor luminescence. One resulting raster scan image with applied forward current (100 μA) is presented in Fig. 1(d). Here, purely electroluminescence (EL) from color centers is detected, i.e. no optical excitation is applied. The emitted light is mainly condensed around the interface of p- and n-layers, but also distinct emitters are found in the i-layer.

These isolated emitters in the i-layer are further investigated. The zoom-in of figure 1(d), indicated by a green rectangle, is shown in Fig. 2(a). Multiple emitters with different brightness are clearly visible. The recorded EL spectrum of the emitter labeled by a white circle in Fig. 2(a) exhibits a maximum around 770 nm, as presented in Fig. 2(c). We perform Hanbury-Brown and Twiss (HBT) type measurements[24] to investigate the single photon emission nature of this EL emitter. The obtained second order correlation of the photon emission statistics $g^{(2)}(\tau)$ at various injection currents are shown in Fig. 3(a), left panel. We confirm the single photon emission *via* $g^{(2)}(0)$ being at low injection current $0.29 \pm 0.04$.

To understand the photo-physics of the investigated emitter, we compare the acquired $g^{(2)}(\tau)$ functions to a theoretical model, describing the emission process of the electrically driven emitter (see also Fig. 3(b) left): The electron can be injected from the conduction band (CB) to the excited state only when the excited state is unoccupied (step 1). After radiative decay to the ground state (step 2) the electron recombines with a hole from the valence band (VB) (step 3), hence the system can now be re-pumped. This process can be interpreted as a closed cycle because both carrier injection occur only when the excited and ground state of a single electron is unoccupied. As we observe no bunching, which typically indicates an shelving process to an intermediate state[25], we consider a two-level system consisting of a ground and excited state. We describe the electrical pumping by a rate $r_{12}$ in analogy to a single photon absorption process, and the radiative decay rate by $r_{21}$. The background corrected[26] $g^{(2)}(\tau)$ curves will then effectively be a two-level system[25] with $g^{(2)}(\tau) = 1 - \exp(-\lambda_1 |\tau|)$ where $\lambda_1 = r_{12} + r_{21}$. Using $r_{12} = \alpha I$, where $I$ is the applied current, one can extrapolate the excited state lifetime, $r_{21}^{-1}$ at $I = 0$ from the fit function as a function of the applied current. (see Fig. 3(c) left). We find $r_{21}^{-1} = 57.5 \pm 4.5$ ns. Note that the shown data represents the found EL emitters which exhibit similar EL spectra and $g^{(2)}(\tau)$.

The observed EL spectra and the excited state life time do not match with known color centers in hexagonal silicon carbide such as silicon vacancies[4,5], divacancies[14,16], or any other new single color centers recently reported, including the D1 center in the confined quantum well near the silicon carbide surface[27], or oxidation related defects[17,19,20]. To investigate further, the same area is raster scanned confocally with 730 nm laser illumination without applying any bias. The same emitter stays dark in PL scan. Instead, many similar emitters can be found. The PL spectrum of a representing one, labeled with black circle in Fig. 2(b), gives rise to a maximum at around 800 nm. In contrast to the EL emitter, the PL detected $g^{(2)}(\tau)$ curves in Fig. 3(a) right panel shows bunching. As already pointed out, these bunching shoulders typically indicate a meta-stable shelving state. Such a behavior is commonly found for atomic scale defects in diamond[7,28] and SiC[4,5] and can lead to spin-dependent photon emission. To model the $g^{(2)}(\tau)$ function we use a three-level model [see the inset in the right panel of Fig. 3(b)][25,29] with,

$$g^{(2)}(\tau) = 1 - (1+a)\exp(-\gamma_1|\tau|) + a\exp(-\gamma_2|\tau|),$$

where $\gamma_1 = r_{12} + r_{21}$, $\gamma_2 = r_{31} + r_{21}r_{12}/\gamma_1$ and $a = r_{12}r_{23}/(\gamma_1 r_{31})$. Similar to the two-level model, $\gamma_1$ depends of the optical excitation power $P_{\text{opt}}$, thus $\gamma_1$ and $\gamma_2$ can be rewritten as:

$$\gamma_1 = r_{21}(1 + \beta P_{\text{opt}}),$$

$$\gamma_2 = r_{31} + r_{23}\beta P_{\text{opt}}/(1 + \beta P_{\text{opt}}).$$

From figure 3(c), we find $r_{21}^{-1} = 4.55 \pm 0.4$ ns, $r_{23}^{-1} = 450 \pm 7$ ns, and $r_{31}^{-1} = 4.54 \pm 0.3\ \mu s$. Saturation behaviors of the photon emission of both EL and PL emitters are also fitted using these rate models, and we find the saturated count rate of 360 kcps and 171 kcps, respectively [see figure 3(d)].

We also investigate the polarization of the emitted single photons. Both emitter types show very well linear polarized dipole pattern as shown in Fig. 4(a) and (b). Note that non-ideal linear polarization could be found from some of the found PL emitters as can be seen in Fig. 4(a). Because bright and inhomogeneous background emission could be found often around such emitter, we ascribe it to imperfect background correction. For the 25 tested PL emitters we find three distinct dipolar orientations, identical in absorption and excitation. The dipole axes are very well matched to the crystal axes of 4H-SiC, which have three-fold symmetry. The color-coded polarization axes for the PL emitters are shown in Fig. 4(c). The polarized photons emitted from the 10 tested EL emitters show only two polarization axes and match with two crystal axes. Based on the experimental finding that the PL and EL defect types vary in spatial localization, and their photons differ in energy and excited state lifetimes as well as polarisation, we conclude that these defects do not have the same origin, despite of being on the same chip. Many new color centers have been found on the SiC surface without ion implantation and electron irradiation[17,19,20,27]. Many of them are attributed to defects in SiC at the SiC/SiO$_2$ interface because the tested SiC samples were oxidized and the optical polarization dependence revealed the three fold symmetry of the hexagonal SiC structures without irradiation of the sample by accelerated particles[17,19]. Because the newly found PL and EL emitters in this report show the same optical polarization dependence, we tentatively relate them to the SiC surface at the oxide interface as well. In order to understand whether the new emitters in this report are artificially created by electron irradiation, we also investigate the non-irradiated p-i-n devices. Figure 1(e) shows the found PL emitters under optical illumination in the non-irradiated sample. Many emitters can be identified as can be seen in the zoom-in in Fig. 2(e).

Their PL spectra are red-shifted [see Fig. 2(f)] compared to the EL and PL emitters found in the irradiated device. Note that PL spectra of these emitters are also not identical to any color centers recently reported[17,19,20,27]. Due to the spectral difference, we conclude, that the PL emitters found in the irradiated device are indeed due to electron irradiation. Due to the limited number of working devices, we could not test if the EL emitters are also observable in the non-irradiated device.

In summary, we found new color centers in both electron irradiated and non-irradiated SiC lateral p-i-n diodes. Among three distinct color centers, one type can be observed only by charge carrier injection into the intrinsic layer, while two other types can be observed by optical excitation in both doped and intrinsic layers. While the exact origin is not known, the surface oxide can be responsible for the origin of the PL emitters found in both irradiated and non-irradiated device. These newly found optically and electrically driven emitters show clear signatures of non-classical light. The respective emission spectra are found within the visible to near-infrared spectral regime. Both emitter types in the irradiated samples exhibit linearly polarized light. The potential further integration into modern electronic devices once more show the promising properties of SiC as a new material platform for quantum applications.

This work was supported by ERC grant SMeL, the Max Planck Society and the Humboldt Foundation, the Swedish Research Council (VR 2016-04068), German Federal Ministry of Education and Research (BMBF) through the ERA.Net RUS Plus Project DIABASE, Carl Trygger Stiftelse för Vetenskaplig Forskning (CTS 15:339), Swedish Energy Agency (43611-1), and KIST institutional programs (Grants No. 2E27231, 2E27110). We thank Roman Kolesov, and Rainer Stöhr for fruitful discussions and experimental aid, Marion Hagel for her help in wire-bonding the diodes, and Alexander Lohrmann for the electrode design. We also acknowledge motivating discussions with Michel Bockstedte and Adam Gali.

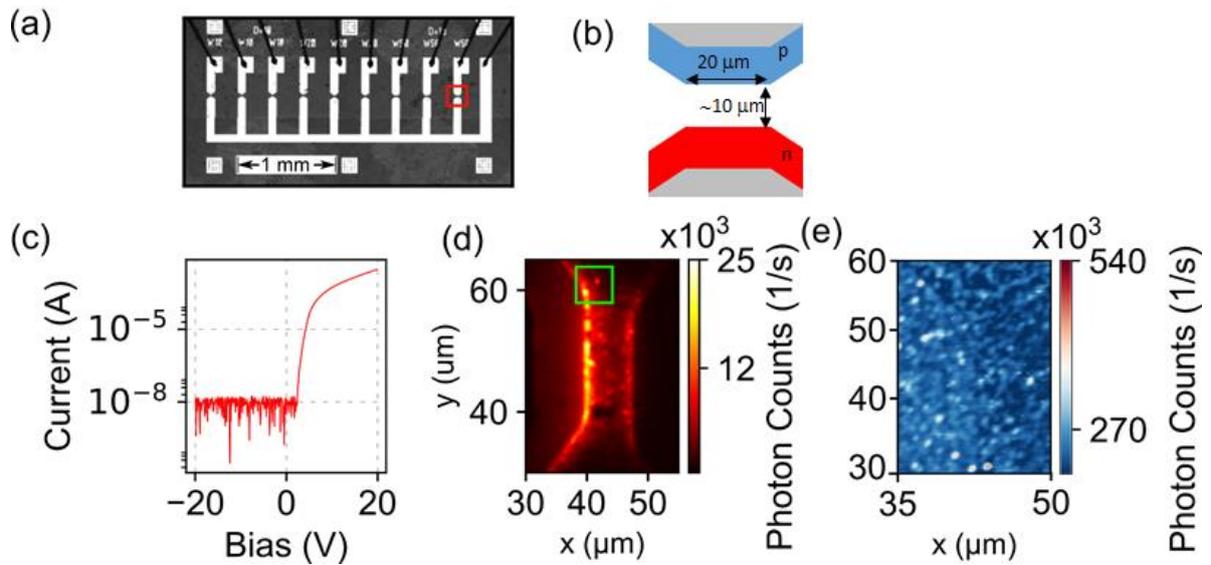

Figure 1. (a) An optical microscopy picture showing the device geometry of the lateral pin diodes. The n- and p-type doped area are covered by metal contacts which are wired bonded for bias application. The red box marked in (a) is the p-i-n junction whose geometry is shown in (b), metal contact is gray colored. (c) An IV-curve of the measured p-i-n-device. The reverse bias current is limited by the sensitivity of the used source-and-measure unit. (d) A confocally scanned map of the biased p-i-n device visualizing the electroluminescence. No optical filter was used. (e) A confocal scanned map of the non-biased p-i-n device surface of the non-irradiated sample, different from (d). 800 nm long-pass filter was used, and 5 mW 638 nm laser was used for excitation.

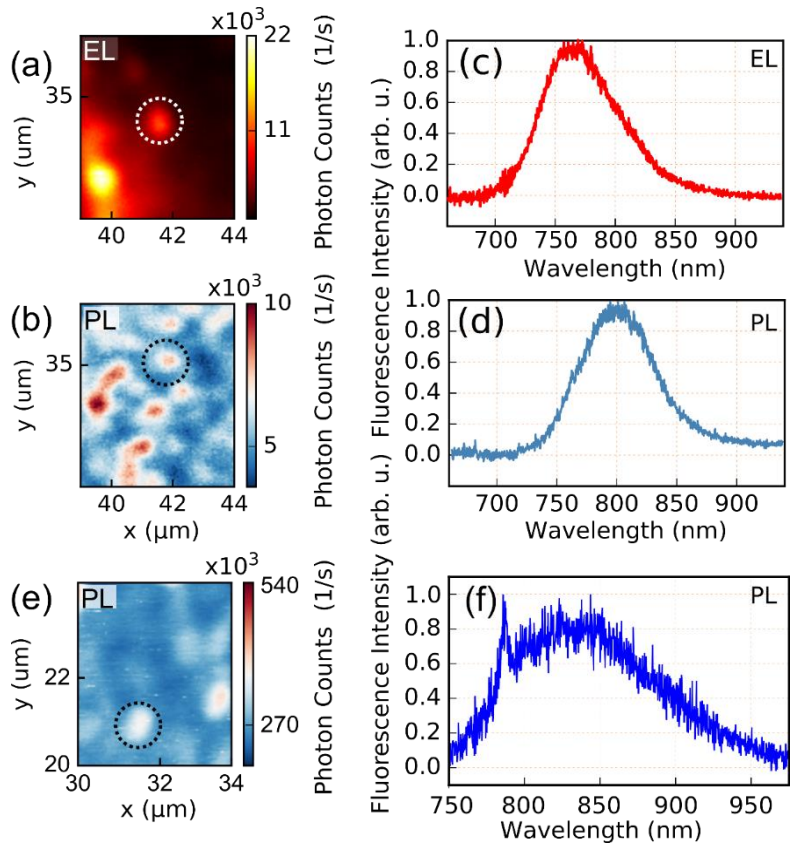

FIG. 2. (a) EL and (b) PL scanned zoom-in of the green area in Fig. 1(d). (c) A luminescence spectrum of the EL emitter in (a). (d) A luminescence spectrum of the PL emitter (different from EL emitter) in (b) excited with 638 nm laser illumination filtered by a 690 nm long-pass. (e) A zoom-in of a randomly selected area of the non-irradiated device in Fig. 1(e). (f) A PL spectrum taken from an emitter in (e). The spectra in (c), (d), and (f) are background corrected.

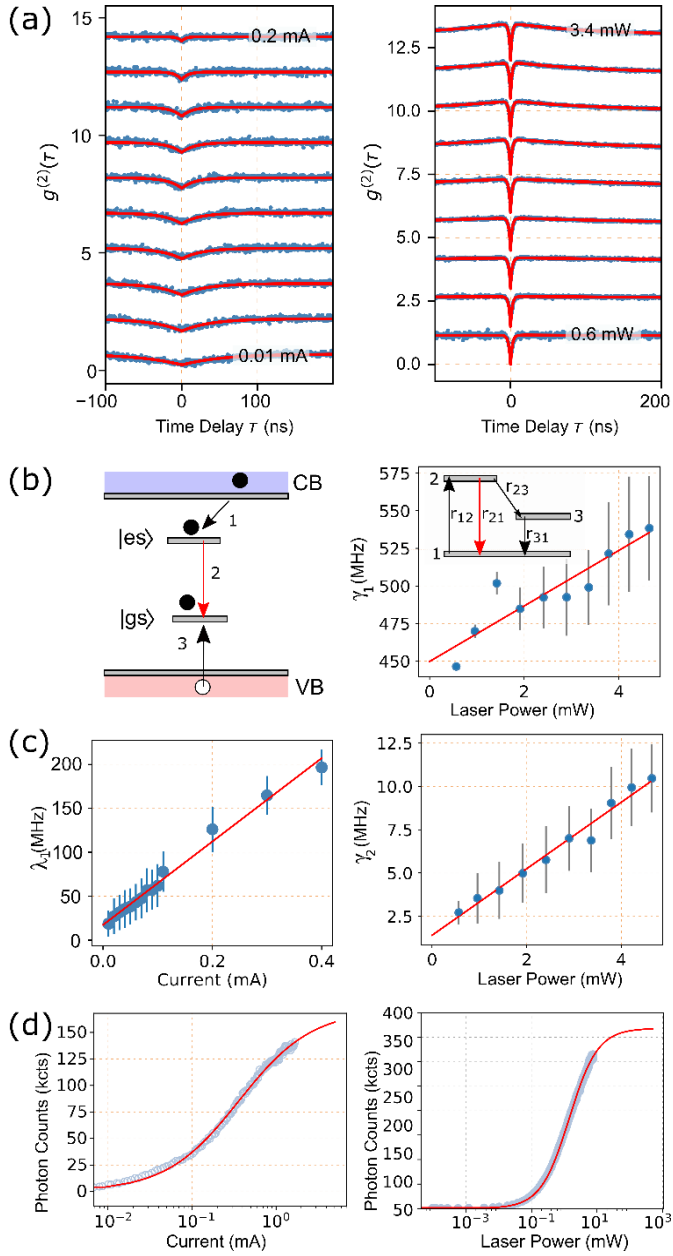

FIG. 3. (a) BG corrected $g^{(2)}(\tau)$ auto-correlations of the EL emitter as a function of the applied current (left), and the PL emitter as a function of the laser power (right). (b) Left: The rate model for the EL emitters. The black and white filled circles represent electrons and holes, respectively. See text for details. Right: the 1st PL rate parameters extracted from right panel of (a) as a function of the laser power and the rate model for the PL emitters as an inset. (c) Left: the EL rate parameters extracted from left panel of (a) are plotted as a function of the applied current. Right: the 2nd PL rate parameters extracted from right panel of (a) as a function of the laser power. (d) Photon counts of the EL emitter as a function of the applied current (left) and the PL emitter as a function of the laser power (right).

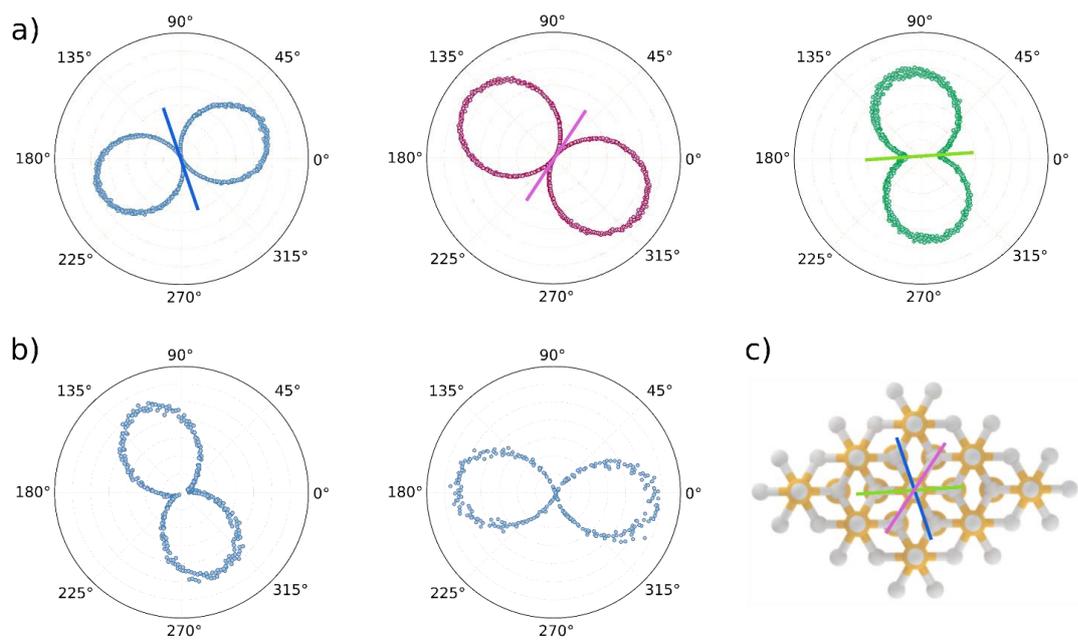

FIG. 4. Photon polarization of (a) three PL emitters and (b) two EL emitters in the irradiated p-i-n device. (c) Color coded dipole axes of the observed PL emitters with respect to the hexagonal lattice.